\documentclass[aps,prl,twocolumn,showpacs,superscriptaddress,groupedaddress]{revtex4}  
\usepackage{graphicx}  
\usepackage{dcolumn}   
\usepackage{bm}        
\usepackage{amssymb}   
\usepackage{amsmath}
\begin{document}
\title{Template for PRL/PRD Papers}
\title{Rounding Effects in Record Statistics}

\author{G.~Wergen} \affiliation{Institute for Theoretical Physics, University of Cologne, 50937 K\"oln, Germany}
\author{D.~Volovik} \affiliation{Center for Polymer Studies and Department of Physics, Boston University, Boston, MA 02215,USA}
\author{S.~Redner} \affiliation{Center for Polymer Studies and Department of Physics, Boston University, Boston, MA 02215,USA}
\author{J.~Krug} \affiliation{Institute for Theoretical Physics, University of Cologne, 50937 K\"oln, Germany}

\date{\today}
\begin{abstract}
  We analyze record-breaking events in time series of continuous random
  variables that are subsequently discretized by rounding to integer
  multiples of a discretization scale $\Delta>0$.  Rounding leads to ties of
  an existing record, thereby reducing the number of new records. For an
  infinite number of random variables that are drawn from distributions with
  a finite upper limit, the number of discrete records is finite, while for
  distributions with a thinner than exponential upper tail, fewer discrete
  records arise compared to continuous variables. In the latter case
  the record sequence becomes highly regular at long times. 
\end{abstract}

\maketitle

\noindent
The statistics of record-breaking events have been widely studied in many
contexts, including sports~\cite{Gembris2002}, evolutionary
biology~\cite{Krug2005}, the theory of spin glasses~\cite{Oliveira2005}, and
the possible role of global warming in the occurrence of record-breaking
temperatures~\cite{Bassett1992,Redner2006,Meehl2009,Wergen2010,Newman2010,Rahmstorf2011}.
Records are defined as the entries in a time series of measurements that
exceed all previous values.  While the record statistics of independent,
identically distributed (iid) random variables (RVs) that are drawn from
continuous distributions are well understood~\cite{Glick1978,Arnold1998}, the
understanding of records drawn from time-dependent
distributions~\cite{Krug2007,Franke2010,Wergen2011} and from series of
correlated RVs~\cite{Majumdar2008,WergenPRE} is still developing.

\begin{figure}
  \includegraphics[width=0.325\textwidth]{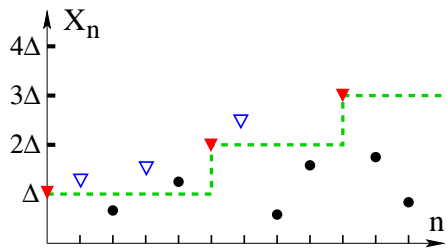}
  \caption{(color online) Effect of rounding down records with discretization
    unit $\Delta$.  Inverted triangles indicate records, with those that
    survive after rounding shown solid.  The dashed line shows the evolution
    of the rounded record value.}
\label{sketch}
\end{figure}

Here we address \emph{discreteness effects} on record statistics.
Conventionally, records are recorded from variables that are drawn from a
continuous distribution.  However, in all practical applications, technical
limitations cause observations to be discrete, even if the underlying
distribution is continuous. In sports or meteorology, distance, time,
temperature, or precipitation measurements are always rounded to a certain
accuracy~\cite{Gembris2002,Meehl2009,Wergen2010}, resulting in an effective
discrete distribution of RVs. Thus ties of existing records can arise, which
alters the probability for a record to occur in any given observation
(Fig.~\ref{sketch}).

For RVs that are explicitly drawn from discrete distributions, the effect of
ties strongly affects the number of
records~\cite{Vervaat1973,Prodinger1996,Gouet2005,Key2005,Gouet2007}.  For
related $\delta$-records and geometric records, where a new record arises
only if the current observation exceeds the current record by a fixed
constant $\delta$~\cite{Gouet2007,Balakrishnan1996} or by a fixed
fraction~\cite{Gouet2012}, intriguing statistical properties of records were
found for the three universality classes of extreme value statistics
(EVS)~\cite{Gumbel1954}.  However, the consequences of measuring
\emph{rounded\/} record values that are drawn from continuous underlying
distributions appears not to have been studied previously.

We consider a set of RVs $X_1,...X_N$ and focus on the probability $P_n
\equiv \textrm{Prob}(X_n > X_1,\ldots,X_{n-1})$ that the $n^{\rm th}$
variable in this series is a record.  We denote $P_n$ as the \textit{record
  rate} and $R_n = \sum_{k=1}^n P_k$ as the \textit{record number}.  For
continuous iid RVs, the universal result is $P_n=\frac{1}{n}$ (see,
e.g.,~\cite{Arnold1998, Glick1978}).  Thus for $n\gg1$, $R_n\approx \ln n +
\gamma$, with $\gamma\approx 0.577...$ the Euler constant.  We assume that.
the RVs $X_i$ are discretized in units of a minimal scale $\Delta$.  That is,
each $X_i$ gets rounded to a value of $X_i^\Delta= k\Delta$.  We may consider
(i) \emph{rounding down}, with $k=\lfloor{X_i/\Delta}\rfloor$ and $\lfloor X
\rfloor$ the floor function, which gives the largest integer smaller than
$X$, or (ii) \emph{rounding to the nearest lattice point}, with
$k=\lfloor{X_i/\Delta + \Delta/2}\rfloor$.  Because asymptotic results do not
depend on the rounding protocol, we will discuss only rounding down.  We
define the \textit{strong} record rate
\begin{eqnarray}
\label{P-def}
P_n^\Delta \equiv \textrm{Prob}\big(X_n^\Delta > X_1^\Delta,\ldots,X_{n-1}^\Delta\big),
\end{eqnarray}
in which ties caused by the discretization are \emph{not counted\/} as new
records.  Thus not only $X_n$, but also the rounded value $X_n^\Delta$ has to
be larger than all previous RVs for a new record to occur
(Fig.~\ref{sketch}).


\smallskip\noindent \textbf{General theory, asymptotic results.} For iid RVs
$X_i$ drawn from a distribution with probability density $f(x)$ and
cumulative distribution $F(x) = \int^x dy\, f(y)$, the record rate is
obtained from $P_n = \int dx\,f(x)F^{n-1}(x)$~\cite{Arnold1998}.  For any
continuous density $f(x)$, this integral gives the universal behavior
mentioned above, $P_n=\frac{1}{n}$.  However, if the measurement $X_i$ is
rounded down to $X_i^\Delta$, the integral for $P_n$ breaks into the sum
\begin{eqnarray}\label{Pnd_general}
  P_n^\Delta & = & \sum_k \left[\int_{k\Delta}^{(k+1)\Delta} \!\! dx\,f(x)\right] F^{n-1}(k\Delta)\,, \nonumber \\
  & = & \sum_k \big[F((k\!+\!1)\Delta) - F(k\Delta)\big] F^{n-1}(k\Delta)\,.
\end{eqnarray}
This gives the strong record rate from continuous RVs that are rounded down
to the closest integer multiple of $\Delta$.  We emphasize that in the
practically more relevant case where record values are rounded either up or
down to the closest integer multiple of $\Delta$, the record rate has the
same statistical properties as those from only rounding down.  We now give
asymptotic results for $P_n^\Delta$ for the three basic classes of EVS
\cite{Gumbel1954}: Weibull (distributions with a finite upper limit), Gumbel
(unbounded upper tail decaying faster than any power law), and Fr\'echet
(power-law upper tail).  Our asymptotic approximations for the discrete
record rate $P_n^\Delta$ for these classes of EVS agree well with numerical
results.

\smallskip\textit{Weibull class:} For illustration, we start with the uniform
distribution: $f(x) = 1$ for $x\in[0,1]$ and $0$ otherwise.  For 
discretization scale $\Delta=\frac{1}{L}$, with integer-valued $L>1$,
Eq.~\eqref{Pnd_general} reduces to:
\begin{equation}
\label{Hn}
P_n^{\Delta} = \sum_{k=1}^{\frac{1}{\Delta}-1}
\Delta\left(k\Delta\right)^{n-1} 
= \Delta^n H_{\frac{1}{\Delta}-1,n-1},
\end{equation}
where $H_{m,n}$ is the $m^{\textrm{th}}$ harmonic number of power $n$.  
At some point in the time series of RVs, a record with a rounded value
$1-\Delta$ occurs; this is necessarily the \emph{last record}.  For a fine
discretization scale, $\Delta\ll1$, the sum in \eqref{Hn} can be replaced by
an integral to give $P_n^{\Delta} \approx
\frac{1}{n}\left(1-\Delta\right)^n$.  Thus for any $\Delta>0$, $P_n^{\Delta}$
no longer decays as $\frac{1}{n}$, but instead approaches zero exponentially
with $n$ --- rounding strongly depresses the asymptotic record rate for the
uniform distribution.

A more general example of the Weibull EVS class is $f(x) = \xi(1-x)^{\xi-1}$,
with $\xi>0$ and $x\in[0,1]$.  
By expanding Eq.~\eqref{Pnd_general} to second order for $\Delta\ll 1$, we
find

\begin{align} \label{P_nd_Kum}
 P_n^{\Delta} & \approx  \int_1^{\frac{1}{\Delta}-1}\! dk
  \left[(1\!-\!k\Delta)^{\xi}
-\left(1\!-\!(k\!+\!1)\Delta\right)^{\xi}\right] \nonumber \\
  &  \qquad \qquad\; \times \left[1-(1\!-\!k\Delta)^{\xi}\right]^{n-1}\,, \nonumber \\
  & \approx \begin{cases}\tfrac{1}{n}\left[1-n\Delta^{\xi}\!-\!\tfrac{\Delta\xi}{2}\,\Gamma
    \big(2\!-\!\tfrac{1}{\xi}\big) n^{{1}/{\xi}}\right], \; & n\Delta^{\xi}\ll 1, \\
    \tfrac{1}{n}\, \exp(-n\Delta^{\xi}), &  n\Delta^{\xi}\gg 1. \end{cases}
\end{align}
Since the underlying distribution has a bounded support, the total number of
records is again finite.  The results in \eqref{P_nd_Kum} reproduce those
found for the uniform distribution.

\begin{figure}
  \includegraphics[width=0.45\textwidth]{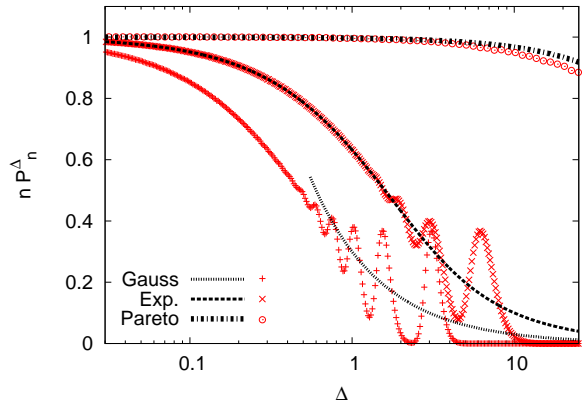}
  \caption{\label{nPn} (color online) Scaled record rate $nP_n^\Delta$ for
    $n=1000$ for the Gaussian, exponential, and Pareto (with $\mu = 1.2$)
    distributions.  Without rounding, $P_n = \frac{1}{n}$.  Simulations
    (symbols) are averaged over $10^6$ time series and over $975\leq n\leq
    1025$ to smooth the data.  Analytical predictions (curves) are shown for
    comparison.  For the origin of the peaks for the Gaussian and exponential
    distributions, see the text following Eq.~\eqref{n-n+}.}
\end{figure}

\smallskip\textit{Gumbel class:} As a basic example, we treat
the exponential distribution $f(x) = e^{-{x}}$.  For
$n\gg1$ we replace the sum in Eq.~\eqref{Pnd_general} by an integral and find
\begin{eqnarray} 
\label{P_nd_exp}
P_n^\Delta \approx \sum_{k=1}^{\infty} e^{-k\Delta} (1\!-\!e^{-k\Delta})^n 
\approx \frac{1}{n\Delta}(1\!-\!e^{-\Delta})
\end{eqnarray}
for arbitrary $\Delta\geq 0$, in agreement with findings for the geometric
distribution in Ref.~\cite{Prodinger1996} and with our simulations
(Fig.~\ref{nPn}).  For $\Delta\ll1$, \eqref{P_nd_exp} reduces to $P_n^\Delta
\approx \frac{1}{n}\left(1-\frac{\Delta}{2}\right)$, while for $\Delta\gg1$,
$P_n^\Delta\approx \frac{1}{n\Delta}$.  In contrast to the Weibull class,
$P_n^\Delta$ asymptotically decays as $\frac{1}{n}$ for arbitrary
$\Delta$.

For the Gaussian distribution $f(x) = \frac{1}{\sqrt{2\pi}}\,
e^{-{x^2}/{2}}$, with unit standard deviation, we find that as
$n\rightarrow\infty$
\begin{align}
  P_n^\Delta & \approx \frac{1}{2}\! \int \! dx
  \left[\textrm{erfc}\left(\frac{k\Delta}{\sqrt{2}}\right)
   \! -\!\textrm{erfc}\left(\frac{(k\!+\!1)\Delta}{\sqrt{2}}\right)\right]\!
  {F(x)^{n\!-\!1}}\,, \nonumber \\
  & \approx  \frac{1}{\Delta} \int \mathrm{d}x\; 
   \frac{1}{\sqrt{2\pi}}\, \frac{1}{x}\, e^{-x^2} \,F(x)^{n-1}.
\end{align}
For $n\rightarrow\infty$ we evaluate this integral by the Laplace method by
expanding the integrand about $x^*=\textrm{ln}(n^2/2\pi)$, where $x^*$ is the
mean value of the $n^{\textrm{th}}$ record.  After some calculation, we
obtain
\begin{eqnarray}
\label{pn-gumbel}
 P_n^\Delta \approx \frac{1}{n\Delta} \left[\sqrt{\ln\left(\frac{n^2}{2\pi}\right)}\right]^{-1}.
\end{eqnarray}
Thus the record rate decays slightly faster than $\frac{1}{n}$
(Fig.~\ref{nPn}). Correspondingly, $R_n^\Delta\propto \Delta^{-1} (\ln
n)^{1/2}$, which diverges weakly as $n\rightarrow\infty$.

\smallskip\textit{Fr\'echet class:} A representative for this class is
the Pareto distribution $f(x) = \mu x^{-\mu-1}$, with $x\!>\!1$ and $\mu\!>\!0$.
Using again Eq.~\eqref{Pnd_general}, the asymptotic record rate $P_n^\Delta$
is
\begin{eqnarray} \label{P_nd_par} P_n^\Delta \approx \tfrac{1}{n}\left[1 -
    \tfrac{\Delta}{2}\,\mu\,\Gamma\Big(2+\tfrac{1}{\mu}\Big)
    n^{-{1}/{\mu}}\right]\,.
\end{eqnarray}
In contrast to the two previous classes, the effect of the rounding is
negligible, as $P_n^\Delta \rightarrow P_n$ for $n \to \infty$ and arbitrary
$\Delta$ (Fig.~\ref{nPn}).

\smallskip\noindent\textbf{Small-$\Delta$ regime.} We now focus on the effects of
rounding when the discretization scale is small ($\Delta\ll 1$) for fixed
$n$.  Here we find a useful analogy between the effect of a linear drift in
RVs \cite{Franke2010} and the effect of rounding, and we adapt methods
developed for the former problem to help elucidate rounding effects.  For
small $\Delta$ the general expression \eqref{Pnd_general} for $P_n^\Delta$
simplifies to
\begin{align}
  P_n^\Delta & =  \sum_k \left[\int_{k\Delta}^{(k+1)\Delta} dx\,f(x)\right] F^{n-1}(k\Delta)\,, \nonumber \\
  & =  \int dx\, f(x) F^{n-1}(\lfloor x\rfloor_\Delta)\,, \nonumber \\
  & \approx  \tfrac{1}{n} - n \int dx\, (x-\lfloor x\rfloor_\Delta) f^2(x) F^{n-2}(x)\,.
\end{align} 
Here $\lfloor x \rfloor_\Delta$ is defined as the largest integer multiple of
$\Delta$ that is smaller than $x$.  Thus, in the second line, $k\Delta=\lfloor
x\rfloor_\Delta$ for $k\Delta\leq x<\left(k+1\right)\Delta$, which obviates
writing the sum.  In the last step, we expand to first order in the quantity
$x-\lfloor x\rfloor_\Delta$ and employ the crude assumption that, on average,
$x-\lfloor x\rfloor_\Delta \approx \frac{\Delta}{2}$ to give
\begin{eqnarray}\label{small_d_general}
 P_n^\Delta \approx \tfrac{1}{n}\left(1-\tfrac{\Delta}{2}n^2 \mathcal{I}_n\right),
\end{eqnarray}
where $\mathcal{I}_n\equiv \int dx\, f^2(x) F^{n-2}(x)$.  The approximation
underlying \eqref{small_d_general} is valid if $n^2\Delta\mathcal{I}_n\ll 1$.
The quantity $\mathcal{I}_n$ appears in record statistics that arise from
continuous RVs with a linear drift \cite{Franke2010}, whose behavior is known
for a wide range of distributions.  In the following we use the results from
\cite{Franke2010} to determine $P_n^{\Delta}$ in the small-$\Delta$ regime.

\textit{Weibull and Fr\'echet classes:} For the distribution $f(x) =
\xi(1-x)^{\xi-1}$ introduced above, the approximation given by
Eq.~\eqref{small_d_general} is useful for $\xi>1$ and we find, for
$n\Delta^{\xi}\ll 1$,
\begin{eqnarray}
  P_n^\Delta \approx \tfrac{1}{n} \left[1-\tfrac{\Delta\xi}{2}\,\Gamma\big(2\!-\!\tfrac{1}{\xi}\big)n^{{1}/{\xi}}\right]\,,
\end{eqnarray}
which, for $n\Delta^{\xi}\ll 1$ and $\xi>1$, agrees with the result derived
from our general approach in
Eq.~\eqref{P_nd_Kum}.  
Similarly, for the Pareto distribution we recover Eq.~\eqref{P_nd_par}.

\begin{figure}
\includegraphics[width=0.45\textwidth]{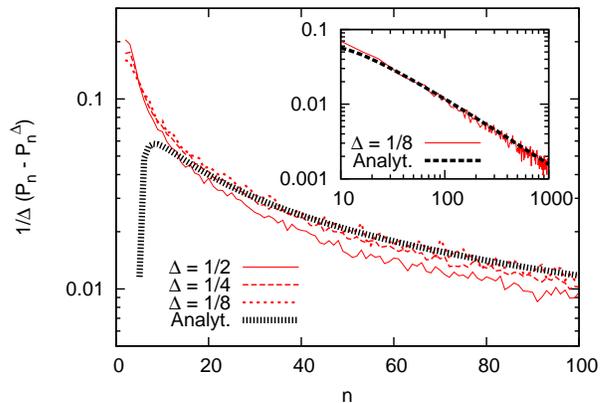}
\caption{\label{Fig:dis_gauss} (color online) Simulations of $P_n^\Delta$ for
  Gaussian RVs in the regime $\sqrt{\ln n}\ll \frac{1}{\Delta}$.  Thin curves
  are $\frac{1}{\Delta}\left(P_n - P_n^\Delta\right)$ for
  $\Delta=\frac{1}{2},\frac{1}{4}$ and $\frac{1}{8}$ and
  $n\in\left[0,100\right]$.  For each $\Delta$, $10^6$ time series were
  simulated.  The thick dashed curve depicts the analytical prediction
  Eq.~\eqref{P_nd_gauss_small_n}. Inset shows the same analysis for
  $\Delta=\frac{1}{8}$ with $n\in\left[1,1000\right]$.}
\end{figure}

\textit{Gumbel class:} For the exponential distribution, we find $P_n^\Delta
\approx \frac{1}{n}\left(1\!-\!\frac{\Delta}{2}\right)$, which agrees with
the small-$\Delta$ behavior of Eq.~\eqref{P_nd_exp}.  For the Gaussian
distribution, the small-$\Delta$ approximation allows us to obtain a new
expression for the record rate when $\sqrt{\ln n} \ll \Delta^{-1}$,
\begin{eqnarray}\label{Gauss_small_d}
P_n^\Delta \approx \frac{1}{n}\left[1-\frac{2\Delta\sqrt{\pi}}{e^2}\sqrt{\textrm{ln}\left(\frac{n^2}{8\pi}\right)}\right]\,.
\end{eqnarray}
The regime $\sqrt{\ln n} \ll \Delta^{-1}$ is not accessible through the
general approach and this range is particularly important for applications, such
as in climatology \cite{Wergen2010}.  For $n\gg 1$ and $\Delta \ll 1$,
Eq.~\eqref{Gauss_small_d} reproduces the numerical simulation values for
$P_n^\Delta$ very accurately (Fig.~\ref{Fig:dis_gauss}).

\smallskip\noindent \textbf{Large-$\Delta$ regime.} For Gumbel-class
distributions that decay at least exponentially fast near the upper limit, we
can provide an alternative description for the record number
$R_n^\Delta$. For these distributions, it is known that the average spacings
between the record events do not increase in time for large
$n$~\cite{Arnold1998}.  Therefore, we may choose a sufficiently large value
of $\Delta$ that almost all records are suppressed because of ties. It then
follows that all discrete values $k\Delta$ (with $k\geq0$) will eventually be
record values and $R_n^\Delta$ is just the sum over the probabilities that a
record has already occurred for a certain value $k\Delta$.  The corresponding
probabilities $\Pi_n\left(k\right)$ for record value $k\Delta$ are given by
$\Pi_n(k) \approx 1 - F(k\Delta)^{n-1}$, which leads to
\begin{eqnarray}\label{new_formula}
  R_n^\Delta \approx \sum_{k=0} \Pi_n\left(k\right) 
  \approx 1 + \sum_{k=1}^{\infty}\left[1-F(k\Delta)^{n-1}\right]\,.
\end{eqnarray}
For elementary Gumbel distributions, interesting properties emerge from
$\Pi_n(k)$.  For a small $n$ and large $k\Delta$, it is obvious that
$\Pi_n(k)\approx 0$.  Conversely, for large $n$ and arbitrary $k\Delta$
eventually $\Pi_n(k)\approx 1$, since $F(k\Delta)<1$ for finite $k\Delta$.

\begin{figure}
\includegraphics[width=0.45\textwidth]{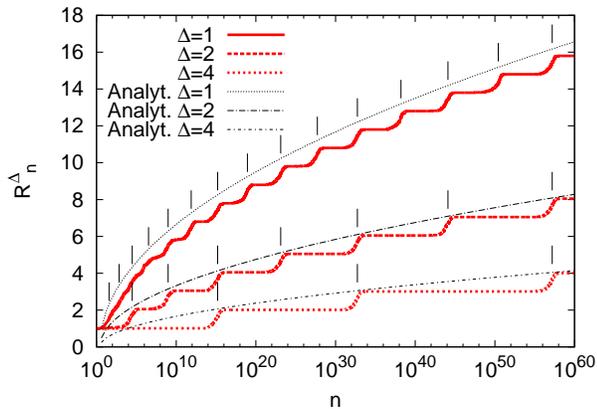}
\caption{\label{steps} (color online) Record number $R_n^\Delta$ for Gaussian
  RVs for $\Delta=1,2,4$.  Data (bold lines) are based on $100$ realizations with a
  maximal $n=10^{60}$. For $n>10^6$ we used an algorithm that directly
  simulates record events by sampling both the distribution and the waiting
  time of the $(k+1)^{\rm st}$ record from the value of the $k^{\rm th}$
  record. Thin lines show the asymptotic behavior predicted by
  Eq.~\eqref{P_nd_gauss_small_n}. The vertical lines show the steps predicted
  by $n\approx\sqrt{2\pi}\,k\,\Delta\, e^{(k\Delta)^2/2}$.}
\end{figure}

We now estimate the regime where $\Pi_n(k)$ switches between $0$ and $1$;
this condition also determines the point where the mean record number
switches from $k-1$ to $k$.  Since $\Pi_n(k)$ will never be exactly $0$ or
$1$, we seek the time $n$, where $\Pi_n(k)$ is either smaller than $\epsilon$
($n\!=\!n_-$) or larger than $1-\epsilon$ ($n\!=\!n_+$) for small
$\epsilon\ll1$.  By elementary means we find
\begin{eqnarray}
\label{n-n+}
  n_- < \frac{\ln\epsilon}{\ln\left[F(k\Delta)\right]}, \qquad 
  n_+ > \frac{\epsilon}{-\ln\left[F(k\Delta)\right]}\,.
\end{eqnarray}
Evidently $\Pi_n(k)$ switches between $0$ and $1$ when $n$ is between $n_-$
and $n_+$, where $n_-$ and $n_+$ are both proportional to
$[\ln\left(F(k\Delta)\right)]^{-1}$.  For the exponential distribution, for
example, we find that $n_- = \epsilon\, e^{k\Delta}$ and $n_+ =
\ln\left(1/\epsilon\right) \,e^{k\Delta}$, so the $k^{\rm th}$ record will
occur at a time proportional to $e^{k\Delta}$, leading to a mean record
number of $R_n^{\Delta} \approx \frac{1}{\Delta} \ln n$.  In the large
$k\Delta$ regime, records occur in an ordered fashion and are well separated
from each other.  The $(k+1)^{\rm st}$ record occurs at time
$e^{\left(k+1\right)\Delta}$, which for $\Delta\gg1$, is much later than the
time of the $k^{\rm th}$ record.  Thus the mean record number undergoes a
step-like periodicity when plotted against $e^n$.  For the Gaussian
distribution, the same approach now predicts that $\Pi_n(k)$ switches for $n
\approx \sqrt{2\pi}k\Delta\; e^{k^2\Delta^2/2}$ (Fig.~\ref{steps}).  For
large $k\Delta$ and large $n$, the mean record number becomes
\begin{eqnarray} \label{P_nd_gauss_small_n} R_n^{\Delta} \approx \sum_{k=0}
  \Pi_n\left(k\right) \approx
  \frac{1}{\Delta}\sqrt{\textrm{ln}\left(\frac{n^2}{2\pi}\right)}~,
\end{eqnarray}
which was already obtained with the general approach above and confirms the
validity of the form for $R_n^\Delta$ given in Eq.~\eqref{new_formula}.  The
step periodicity in $R_n^\Delta$ is the source of the observed peaks
(Fig.~\ref{nPn}) in the record rate $P_n^\Delta$ as a function of $\Delta$
for exponential and Gaussian distributions.

\smallskip\noindent \textbf{Conclusions.} We determined how rounding down
continuous random variables affects the statistics of records.  Our results
directly apply to the practical situation where continuous variables are
rounded either up or down to the closest integer multiple of a fixed
discretization scale $\Delta$.

For distributions with bounded support, rounding leads to an exponential
decay of the record rate, $P_n^\Delta$, and an asymptotically finite record
number. In contrast, for power-law distributions, the effect of rounding
becomes negligible for $n\rightarrow\infty$ and $P_n^{\Delta} \rightarrow
\frac{1}{n}$ independent of $\Delta$.  In the intermediate Gumbel class, the
behavior is more subtle.  For the exponential distribution, $P_n^{\Delta}$
decays as $\frac{1}{n}$ with a $\Delta$-dependent prefactor, while for the
general distribution $f(x) \propto \exp(-|x|^{\beta})$ with $\beta>1$, the
record rate decays as $n^{-1}\, \textrm{ln}\left(n\right)^{1/\beta-1}$.

For underlying distributions that decay at least exponentially, the record
sequence becomes ordered at long times, in marked contrast to independent
record events from continuous iid RVs~\cite{Glick1978,Arnold1998}. While
correlations between record events have been previously observed for RVs that
are drawn from drifting \cite{Wergen2011} or broadening \cite{Krug2007}
distributions, the effect of rounding is much stronger and renders record
events predictable on a time scale that grows exponentially (or faster) with
record number.


To illustrate that rounding effects have an observationally
significant influence on records, we analyzed 50 years of daily temperatures 
from 361 U.S. weather stations \cite{USHCN} along the lines of \cite{Wergen2010}. 
The measurements were reported in integer units of $\Delta = 1^\circ$F
and we considered all 361 $\times$ 365 time series for the individual calendar 
days with an average standard deviation of $\sigma\approx 8.9^\circ$F.
Only $75\%$ of the weak upper (ties allowed) and $78\%$ of 
the weak lower records were also strong records (no ties), in good agreement
with the value of $79\%$ predicted by our analytical result in
Eq.~\eqref{Gauss_small_d}. In this example the effect of ties
on the record rate has a similar magnitude as that of the small warming trend
in the data (cf.~\cite{Redner2006,Meehl2009,Wergen2010}). Thus rounding effects 
should be carefully accounted for if one wishes to use record statistics to detect 
secular trends in data, such as global warming.


GW acknowledges financial support from Friedrich-Ebert-Stiftung and BCGS as
well as the kind hospitality of the Center for Polymer Studies in the early
stages of this work.  DV and SR thank NSF grant DMR-0906504 for partial
financial support of this research. We thank O. Pulkkinen for making us aware of Refs. \cite{Vervaat1973,Prodinger1996}.

\end{document}